\documentclass[namedreferences,hyperref,optionalrh]{spr-sola}
\usepackage{graphicx}        
\usepackage{color}           
\usepackage{xcolor}           
\usepackage{ulem}

\chardef\us=`\_
\newcommand{\kms}{km s$^{-1}$}

\begin{document}

\begin{frontmatter}
\title{Supersonic flows observed by THEMIS related to a coronal bright point and filament}

\author[addressref={aff1},corref,email={garimakarki31@gmail.com}]{\inits{G.}\fnm{Garima}~\snm{Karki}\orcid{0009-0003-3193-7496}}
\author[addressref={aff2,aff3,aff4,aff5},corref]
{\inits{B.}\fnm{Brigitte}~\lnm{Schmieder} \orcid{0000-0003-3364-9183}}
\author[addressref={aff1},corref]{\inits{R.}\fnm{Ramesh}~\lnm{Chandra} \orcid{0000-0002-3518-5856}}
\author[addressref={aff6,aff7},corref]
{\inits{P.}\fnm{Pooja}~\lnm{Devi} \orcid{0000-0003-0713-0329}}
\author[addressref={aff2},corref]
{\inits{B.}\fnm{Pascal}~\lnm{D\'emoulin} \orcid{0000-0001-8215-6532}}
\author[addressref={aff3,aff8},corref]
{\inits{S.}\fnm{Stefaan}~\lnm{Poedts} \orcid{0000-0002-1743-0651}}
\address[id=aff1]{Department of Physics, DSB Campus, Kumaun University, Nainital -- 263001, India}
\address[id=aff2]{LIRA, Observatoire de Paris, Universit\'e PSL, CNRS, Sorbonne Universit\'e, Universit\'e Paris Cité, 5 place Jules Janssen, 92195 Meudon, France} 
\address[id=aff3]{Centre for mathematical Plasma Astrophysics, Dept. of Mathematics, KU Leuven, 3001 Leuven, Belgium}
\address[id=aff4]{SUPA, School of Physics \& Astronomy, University of Glasgow, Glasgow G12 8QQ, UK}
\address[id=aff5]{LUNEX EMMESI Institut, SBIC, Kapteyn straat 1, Noordwijk2201 BB Netherlands}
\address[id=aff6]{Rosseland Centre for Solar Physics, University of Oslo, P.O. Box 1029, Blindern, N-0315 Oslo, Norway}
\address[id=aff7]{Institute of Theoretical Astrophysics, University of Oslo, P.O. Box 1029, Blindern, N-0315 Oslo, Norway}
\address[id=aff8]{Institute of Physics, University of Maria Curie-Skłodowska, Lublin, Poland}

\runningauthor{Karki et al.}
\runningtitle{Supersonic flows observed by THEMIS related to a coronal bright point and filament}

\begin{abstract}
In this paper, we report on the dynamics of the fine structure of a solar quiescent filament observed on September 28, 2023, with the T\'elescope H\'eliographique pour l’Etude du Magn\'etisme et des Instabilit\'es Solaires (THEMIS).
The main aim is to understand the relationship between the supersonic downflows measured in H$\alpha$ at the filament end and an associated coronal bright point. 
We use a cloud-model method to derive the supersonic velocity of the falling, elongated cool blob. Besides, we use H$\alpha$ Global Oscillation Network Group (GONG) data to track the plasma along the filament. 
Repetitive plasma motions are observed along the northern end of the filament. During one event of plasma motion, the THEMIS field of view was centred on the filament end, where supersonic downflows of 89.8 \kms\ were measured with a standard deviation of $\pm$ 0.25 \kms. 
We suggest that the plasma moving along the filament could be falling towards the chromosphere. A ballistic trajectory could confirm this first scenario. However, we could not rule out the second scenario, in which coronal rain forms due to the thermal instability of coronal plasma. In the hot AIA channels, we identify a bright point at the same location, with a temperature reaching about 6 MK. 
In addition, we confirm a counter-streaming flow pattern along the fine filament strands, with widths of less than an arc second, as measured by the high-spatial- and spectral-resolution spectra of THEMIS.
\end{abstract} 

\end{frontmatter}

\section{Introduction}
\label{Intro}
Solar filaments in the corona have been observed for 150 years by many observers, starting with \citet{Sechi1876}. They are fascinating structures when observed at the limb, and many reviews describe their structure, formation, and eruptions; see the reviews by \citet{Tandberg1995, MacKay2010, Labrosse2010} and, more recently, those by \citet{Keppens2025, Liakh2025}, and \citet{Zhou2025}. Solar filaments are elongated, cool, dense plasma structures supported against the Sun's gravity by magnetic fields. Magnetic field lines shape them in the solar corona, often extending down toward the chromosphere and lying above the polarity inversion lines (PILs). 

The dynamics of prominences are fully investigated during their eruptive phases, which lead to coronal mass ejections \citep{Chandra2021, Devi2021, Devi2022, Xing2024}. However, the dynamics of quiescent prominences remain debated. The dynamics of fine structures in filaments remain poorly understood. Filaments consist of many fine, thread-like structures. These fine structures exhibit various dynamics. One example is counter-streaming flows. Counter-streaming flows can exist in the same observed threads as well as in adjacent threads. These flows are in opposite directions. They can be observed both in the plane of the sky and along the line of sight (LOS). These oppositely directed flows were first defined as counter-streaming flows by \citet{Zirker1998}, in which blue- and red-shifted patterns were observed in H$\alpha$ wing images.  They deduced from their observations that the flows were vertical. Various mechanisms have been proposed to explain counter-streaming flows. \citet{Panesar2020} observed counter-streaming flows in an intermediate filament within the enhanced network and proposed that these flows result from the eruption of small-scale jets at both ends of the filament. Counter-streaming mass flows in on-disk filaments and limb prominences can be driven by repeated coronal jets originating at one of the footpoints of the filament or prominence, as demonstrated by \cite{ZhouC2025} using observational results and 3D Magnetohydrodynamics (MHD) numerical simulations. \citet{karki2025} discovered, in high-resolution H$\alpha$ spectra, very tiny threads with arcsecond diameters, driven by counter-streaming flows reaching 20 \kms. These supersonic flows could not be explained by a mostly horizontal velocity component, since the filament was at the disk centre, with no possible projection effects.  Kink waves were suggested to be relevant to these high velocities. Such high Doppler shifts were also observed in a filament-in-formation using the Multi-Subtractive Double-Pass Spectrograph (MSDP) on T\'elescope H\'eliographique pour l’Etude du Magn\'etisme et des Instabilit\'es Solaires \citep[THEMIS;][]{Schmieder2014}. The blue- and red-shifted threads were visible in H$\alpha$ at $\pm$ 0.45 \AA\ near a footpoint, and the Doppler shifts were estimated using cloud-model techniques \citep{Beckers1964} to be in excess of 10 \kms. A question arises: is counter-streaming a precondition for the eruption of a filament, as suggested by measurements of counter-streaming in an erupting filament \citep{Schmieder2008}?

Filament dynamics must account for oscillations \citep{Arregui2018}. They can be produced by waves or due to bulk flows. Small- or large-amplitude oscillations (SAO/LAO) are frequently detected \citep{Luna2017,Luna2018, Devi2022adspr, Joshi2023}. Their periods are on the order of an hour, and the amplitude is less than 10 \kms\ for SAO or greater than 10 \kms\ for LAO.

The LAOs may correspond to counter-streamings with oscillating threads around magnetic dips \citep{Chen2014, Zhou2020}. It preferentially exists in a highly twisted flux rope. In a weak twist flux rope, the counter-streaming consists of alternating unidirectional flows \citep{Zhou2017}. Both types of threads exist in prominences, and the counter-streaming of prominences might consist of both longitudinal oscillations and unidirectional flows, with the proportions determined by the twist of the supporting flux rope. \citet{Guo2022_prom} investigates the types of thread flows in magnetic flux ropes with different numbers of twists using high-resolution pseudo-3D simulations. In the weakly twisted flux rope, the majority of threads are short-lived, dynamic threads (83\%) that do not require magnetic dips and form high-speed flows within the filament. 

Supersonic flows could arise from thermal instability in the corona. It is relatively common to observe cool plasma within the hot corona, with a temperature difference of two orders of magnitude. It corresponds to prominence plasma or coronal rain. Both are well observed on the limb. Prominences are easy to observe on the disk as filaments. Coronal rain is rare and difficult to observe on the disk \citep{Antolin2012}. \cite{Li2021} reported coronal condensation and rain on the solar disk using images from the Solar TErrestrial RElations Observatory (STEREO) - Solar Dynamics Observatory (SDO) quadrature. STEREO well observes the coronal rain on the limb. In SDO on-disk observations, condensations appeared as dark features in Atmospheric Imaging Assembly (AIA) 304 \AA\ images and moved downward as on-disk coronal rain.

Two physical mechanisms are proposed to explain the cooling in coronal plasma that produces coronal rain: one is similar to the formation of prominence plasma in coronal loops, where heating at one footpoint increases plasma pressure and drives siphon flow. This flow may trigger thermal instability, leading to the formation of cool plasma \citep{Froment2020, Keppens2025, Zhou2025}.  In the second case, the model proposes filament formation through the injection of chromospheric plasma \citep{Wang2018, Zhou2025Review}. 

This study focuses on the dynamics of the fine structures of a quiescent filament using spectroscopic data obtained with THEMIS \citep{Schmieder2025}. This filament was the target of the THEMIS-Interface Region Imaging Spectrograph \citep[IRIS;][]{DePontieu2014} campaign in September 2023. The high spatial resolution of the H$\alpha$ spectra from THEMIS enables us to analyse the Doppler shift of each strand that forms the filament core. A counter-streaming flow was observed in some parts of the filament on September 29 \citep{karki2025}. The formation of the filament's feet, its extension, and its 3D magnetic reconstruction were studied using Hinode and Helioseismic and Magnetic Imager (HMI) \citep{Karki2026}. 

In this paper, we analyse the H$\alpha$ spectra of the same filament, observed the day before with THEMIS (September 28). Surprisingly, we measure supersonic downflows at one end of the filament. At the same location, a bright point (BP) is detected in the AIA filters. Could it be the downflow plasma identified as coronal rain, or does it belong to the filament?
The paper is structured as follows: Section~\ref{Instrument} presents the instruments, datasets, and analytical techniques. The filament imaging and spectral observations (counter-streaming) are included in Section~\ref{obs}. In particular, Section~\ref{downflow} describes the observed high downflows at the edge of the filament channel. Blobs flowing along the filament are described in Section~\ref{plasmoid}, and their relationship to the hot EUV BP observed simultaneously at the same location is presented in Section~\ref{EUV}. Section~\ref{Conclusion} presents the discussion and conclusion of the study.

\begin{figure}    
\centerline{\includegraphics[width=1.\textwidth,clip=]{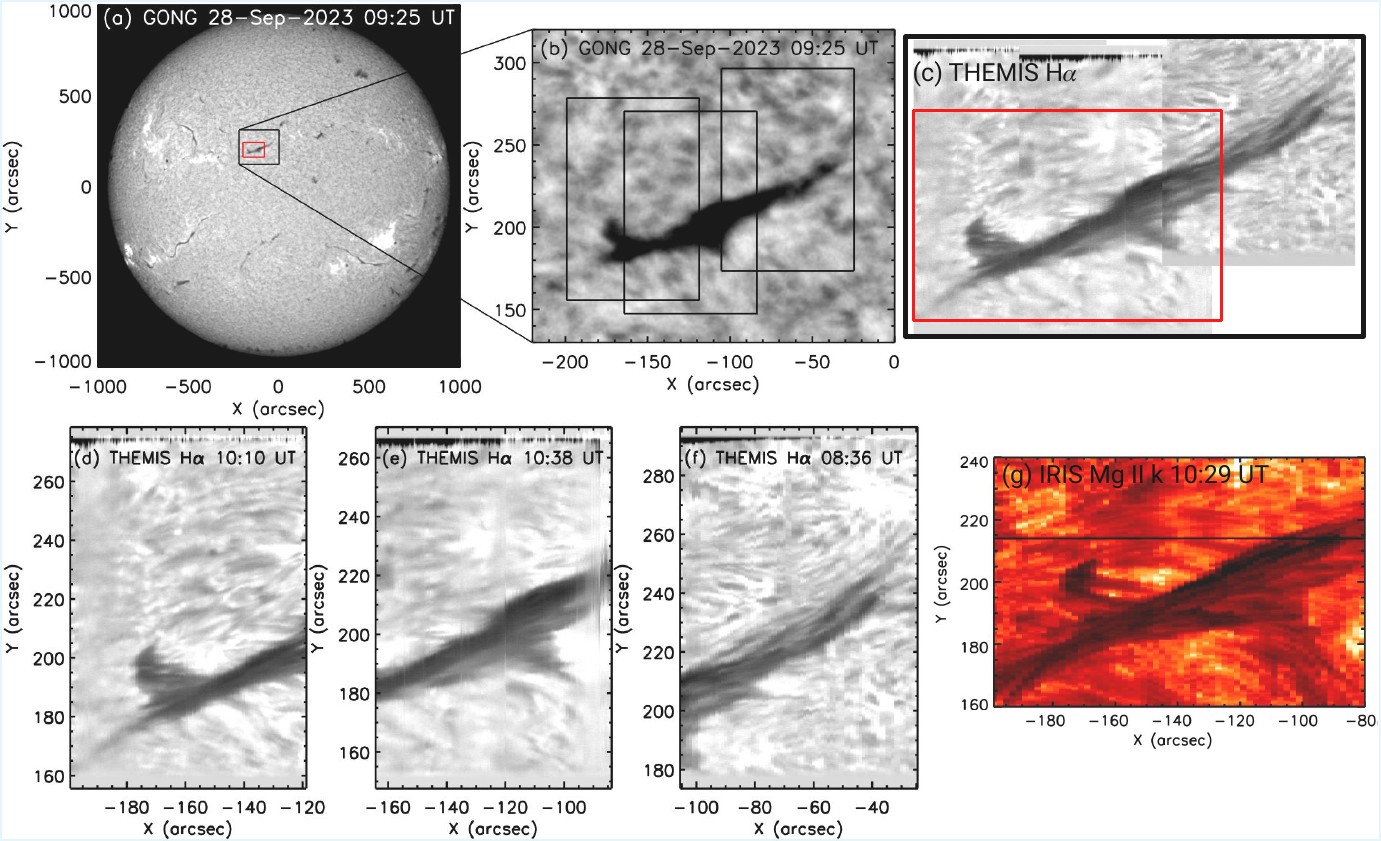}}
       \caption{Solar filament observed by GONG 
       on September 28, 2023 (panels a and b with a zoom image) and THEMIS (panel c). The red box in a and c indicates the IRIS FOV for context. In panel b, the three black boxes represent the FOVs of three sections of the filament observed by THEMIS, from left to right: east, centre, west (see Table~\ref{tab:THEMIS}). The THEMIS image in panel c is reconstructed by combining the three different sections of the filament observed by THEMIS at 08:36 UT, 10:10 UT and 10:38 UT on September 28, 2023 (panels d--f). Panel g shows the filament section observed by IRIS for context. THEMIS and IRIS images are reconstructed using spectra at the line centres of H$\alpha$ and Mg II k, respectively.}
\label{gong}
\end{figure}

\section{Instruments and data sets}
\label{Instrument}
To study the dynamics and fine structures of a quiescent filament on September 28, 2023, we make use of data from the following space- and ground-based observatories. This filament was the target of the THEMIS-IRIS campaign in September 2023.

THEMIS/``Multi Raies'' (MTR) mode can observe multiple lines, e.g., H$\alpha$ (6562.8 \AA), He I (5876 \AA\- He I  D3). THEMIS scans a small portion of the Sun ($ 80'' \times 120''$) with a pixel size of 0.06$''$ along the slit and a spectral dispersion of $\approx$ 3.067 m\AA\ per pixel.  This mode automatically ensures perfect co-alignment of the maps across different wavelengths. Here, we focus on the H$\alpha$ observations with a passband centred at 6563 \AA\ (6.3 \AA\ wide). The slit step is either 0.5$''$ or 1$''$, with a slit width of 0.5$''$. Depending on the targets and viewing conditions, the H$\alpha$ exposure time ranges from 0.05 to 0.2 s. The details of the observations taken by THEMIS on September 28, 2023  between 08:23 UT and 11:19 UT are presented in Table~\ref{tab:THEMIS} (see the Appendix). Here, we focus mainly on the northwestern part of the filament, observed during three raster scans at 08:36, 08:37, and 08:39 UT. At 08:36 UT, we observe the maximum downflow. To characterise the downflow and the filament dynamics, including counter-streaming flows, we use raster scans at 08:36 and 08:37 UT. 

For the morphological evolution of the filament channel, we used data from AIA \citep{Lemen2012} onboard SDO \citep{Pesnell2012}. We have primarily used observations with a pixel scale of 0.6$''$ per pixel and a temporal resolution of 12 s in the AIA 94, 171, 193, 211, and 304 \AA\ filters. HMI magnetograms with a pixel size of $0.5''$ are used to investigate the magnetic configuration of the BP region.

IRIS observed the filament in coordination with THEMIS with its field of view (FOV) centred on the central section of THEMIS. However, the FOV does not cover the region of interest: the west end of the filament with high flows.

Solar filaments are observed as dark features on the solar disk in H$\alpha$ images due to the absorption of chromospheric radiation. For full-Sun H$\alpha$ observations of the filament, we use data from the Global Oscillation Network Group \citep[GONG;][]{Harvey1996}, which provides continuous full-Sun observations with a pixel size of 1$''$ and a temporal resolution of 1 min. 

Observations from all of the above instruments are aligned on September 28, 2023 at 09:00 UT to correct for solar differential rotation using the $drot\_map$ routine in SolarSoft (SSW) for AIA, HMI, and GONG observations. The alignment of the data from these three instruments has been checked by cross-correlation, and the error in their co-alignment with THEMIS is estimated at 1$''$, the GONG pixel size.

\begin{figure}[!t]    
\centerline{\includegraphics[width=1.0\textwidth,clip=]{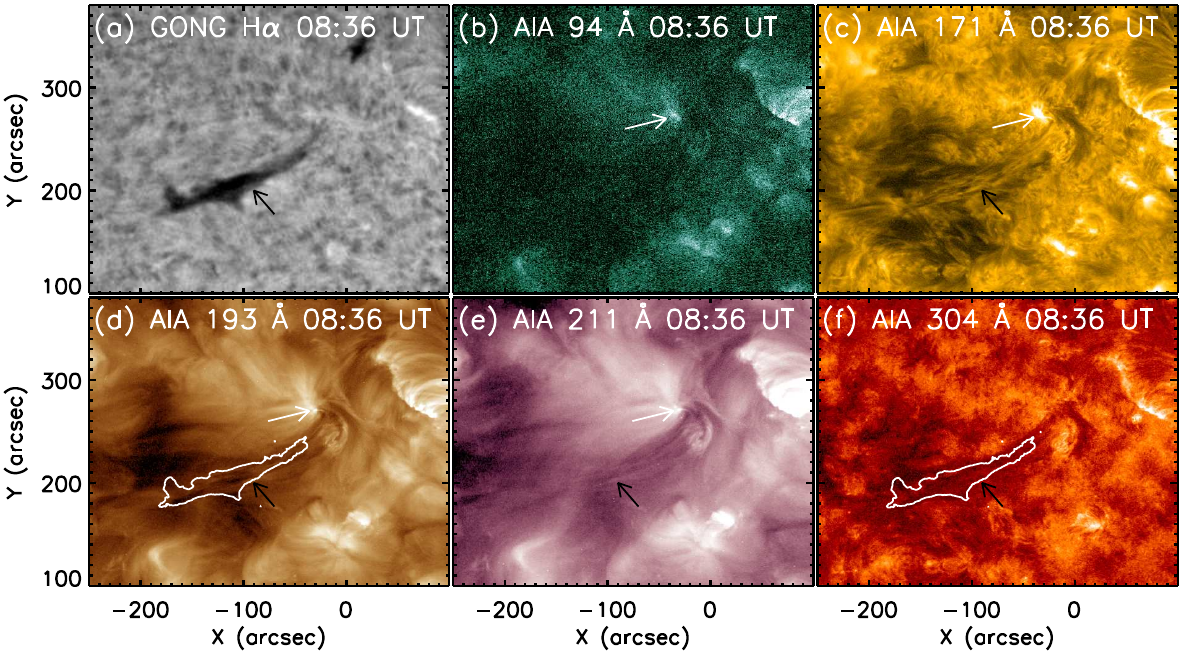}}
     \caption{Filament observed in H$\alpha$ with GONG (panel a) and in different AIA channels (panels b-f). The filament in AIA 171, 193 and 211 \AA\ appears as dark structures {indicated by black arrows}. A BP, indicated by a white arrow, is present in AIA 94, 171, 193 and 211 \AA\ but absent in AIA 304 \AA. White contours in panels d and f are the filament contours observed in H$\alpha$ with GONG. An animation of this figure is attached.}
\label{AIA_BP}
\end{figure}

\section{Filament observations}
\label{obs}
\subsection{Morphology}
A quiescent filament on September 28, 2023, was located on the solar disk at N17 E07. THEMIS targeted the filament in coordination with IRIS. We studied the magnetic evolution of the filament region using the HMI and Hinode/SOT instruments in a previous paper \citep{Karki2026}.  The 3D MHD reconstruction of the filament reveals that its magnetic configuration evolves into a full twisted flux rope the next day.

\begin{figure}    
\centerline{\includegraphics[width=1\textwidth,clip=]{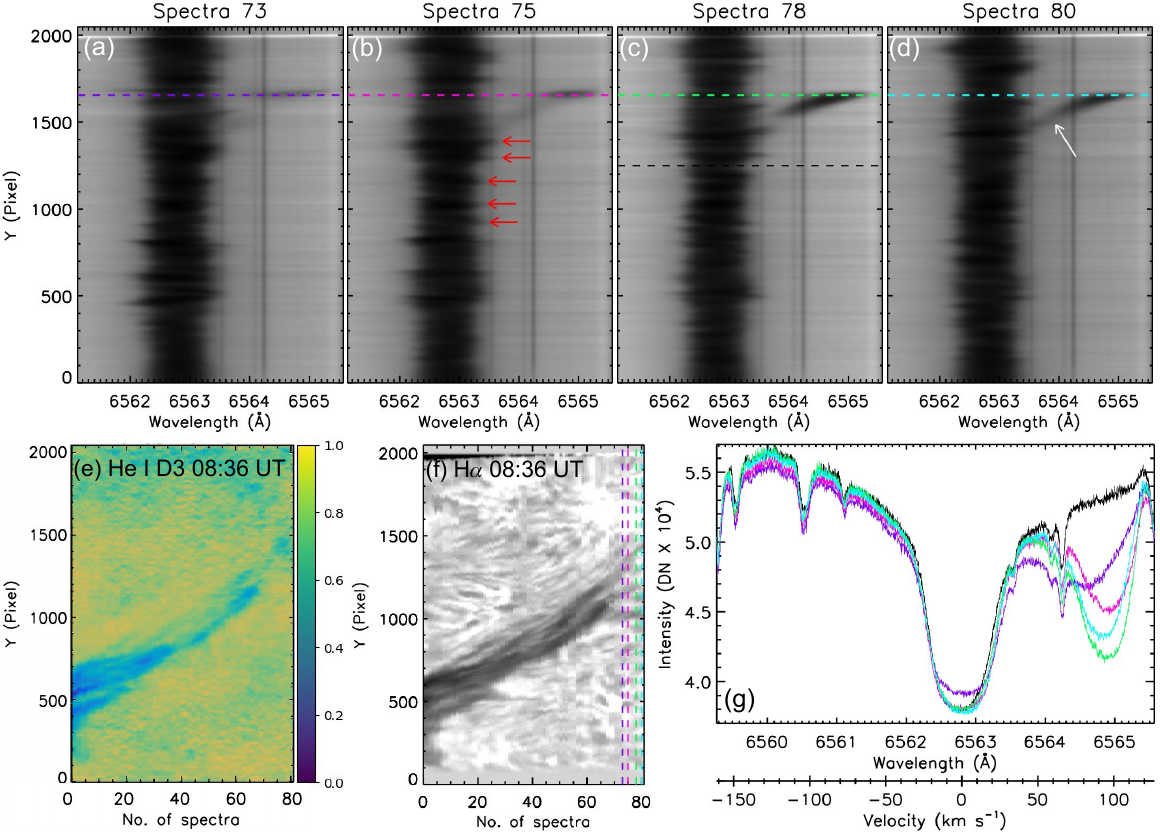}}
\caption{H$\alpha$ spectra (73, 75, 78 and 80) of the filament west end during the scan obtained at 08:36 UT (panels a, b, c, and d). The $x$-axis is the wavelength, and the $y$-axis is along the slit in units of pixels. The pixel size is around 0.06$''$ (45 km). The scan consists of 81 spectra with a step of 1$''$. Large plasma flows are observed in the red wing of H$\alpha$ ($y$ in the range [1500, 1700]). The white arrow in spectrum 80 indicates the progressive extension of the H$\alpha$ line towards the large-flow region in this $y$-domain. Red arrows in spectrum 75 indicate the spectra of the finger-shaped end of the filament. Panels e and f show the filament observed with THEMIS in He I D$_3$ wavelength in normalised values and in the H$\alpha$ line centre (Intensity between 3.8 and 5.6 DN $\times 10^4$), respectively. The respective positions of the spectra are overlaid 
in panel f, with violet (spectra 73), magenta (spectra 75), green (spectra 78), and cyan (spectra 80) dashed vertical lines. The same colour convention is used for the dashed horizontal lines in panels a, b, c, and d to indicate the $y$-position (at 1655 pixels) where the H$\alpha$ profiles are drawn in panel g as continuous lines of the respective colours. The black profile in panel g is the reference profile, corresponding to the black dashed line in spectrum 78 (panel c) in the chromosphere. An animation of these panels a-d and f is attached to this figure.} 
\label{rain}
\end{figure}

Figure~\ref{gong} shows the filament observed by GONG on September 28. The red box in panel a represents the IRIS FOV, while the black box indicates the FOV in panel b. THEMIS observed different parts of the filament, as described in Table~\ref{tab:THEMIS} of the Appendix. The FOVs of THEMIS, which observe different sections of the filament, are indicated by three black boxes in Figure~\ref{gong}b. Panels d, e, and f of Figure~\ref{gong} show different sections of the filament observed by THEMIS. Filament fine structures are clearly visible in the high-spatial-resolution THEMIS-reconstructed images and in the IRIS map (panel g). The complete filament obtained by combining THEMIS observations of different parts of the filament shows an ensemble of fine threads elongated along the filament axis and making a small angle with this axis ($\approx$ 20 degrees, Figure~\ref{gong}c). 
The filament is also visible in He I D3 and shows similar fine structures (Figure~\ref{rain}e). This was possible because of THEMIS's high spectral resolution. This is remarkable since filaments on the disk are very rarely observed in He I. In contrast, prominences are frequently observed in He I D3 \citep{Koza2017}, and their multiplet characteristics favour magnetic-field computation \citep{Levens2016}.  

The filament observed in H$\alpha$ (Figure \ref{AIA_BP}a) is embedded in a large filament channel visible in AIA EUV wavebands, clearly seen in AIA 304 \AA\ (Figure~\ref{AIA_BP}f). The denser part of the filament in AIA 171, 193, and 211 \AA\ appears darker {(Figure~\ref{AIA_BP}c--e)} due to the similar optical thickness of EUV wavelengths with H$\alpha$ \citep{Anzer2008, Heinzel2008}. A BP close to the filament is observed in hot channels of AIA (Figure \ref{AIA_BP}b--e).  We discuss these EUV observations in Section \ref{EUV}.

\begin{figure}    
\centerline{\includegraphics[width=1.\textwidth,clip=]{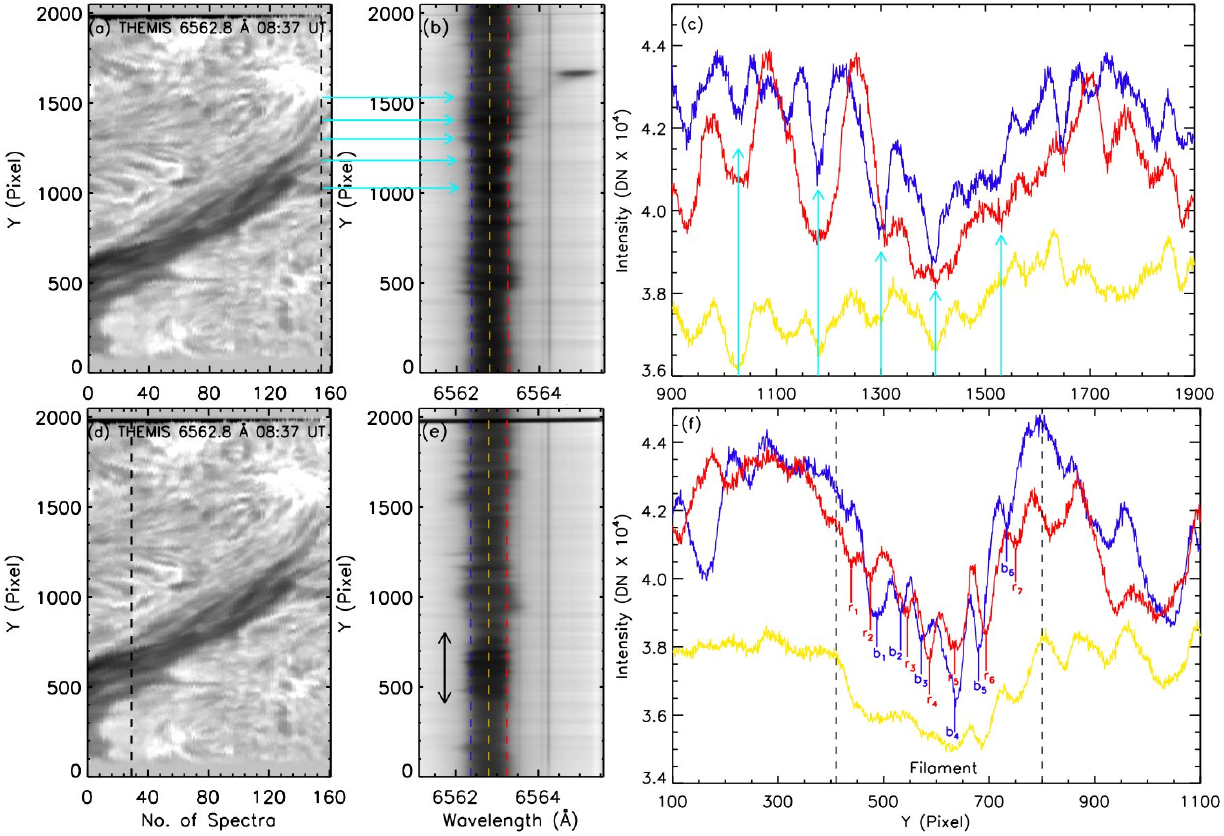}}
\caption{
Filament maps in the H$\alpha$ line centre obtained at 08:37 UT (panels a and d); a vertical dashed black line indicates the position of the slit used to obtain the spectra shown in panels~b and e, respectively. In panels a and d, the $x-$axis indicates the spectrum number; the step between spectra is 0.5$''$; and the $y$ axis indicates the pixel number (pixel size is 0.06$'' \approx$ 45 km). 
In panels b and e, three vertical lines are added: in yellow at the H$\alpha$ line centre, and in red/blue at the red/blue wing at $\lambda$ equal to the H$\alpha$ line centre $\pm$ 0.44 \AA. 
The double black arrow in panel~e indicates the position of the filament in this spectrum. Panels~c and f show the intensity curves in DN units along the three vertical lines drawn in panels b and e, respectively. 
The cyan arrows indicate the position of the fingers in panels b and c. In panel c, the red and blue dips (at the finger locations) are at the same position along the slit.  
In contrast, in panel f, we observe multiple blue and red dips that are not at the same position along the slit, indicating LOS counter-streaming at speeds of $\pm$ 20 \kms.}
\label{Doppler28}
\end{figure}

\subsection{Supersonic flows}
\label{downflow}
THEMIS observed different sections of the filament using multiple rasters on September 28, 2023, as shown in Table~\ref{tab:THEMIS}. Some of these rasters, at 08:36 UT and 08:37 UT, were obtained in the early morning with very good seeing (low atmospheric turbulence and low humidity). The first raster, done at 08:36 UT with an $x$ step of 1$''$, consists of 81 spectra, while the raster at 08:37 UT, done with an $x$ step of 0.5$''$, consists of 161 spectra (Figures \ref{rain} and \ref{Doppler28}, respectively). The very wide spectral window (6 \AA) of the present datasets allows us to detect higher-velocity Doppler components than instruments that produce images at discrete wavelengths within a restricted wavelength domain (about 2 \AA), such as the SST. The THEMIS rasters show large plasma flows in the red wing of H$\alpha$, visible up to + 2.5 \AA\ (Figure~\ref{rain}). These flows were observed 
at 08:36 UT and 08:37 UT; they were no longer visible in the next observation at 08:39 UT. Besides, in the mode of rastering a region, we benefit from the high spectral resolution of THEMIS: 2048 pixels covering approximately  6.3 \AA. The pixel resolution is 3.067 m\AA. The spectral resolution is a factor of 100 higher than instruments that image in different wavelengths (e.g., SST).

We analyse the H$\alpha$ spectra of the filament on September 28 from these two raster scans. The movie of the H$\alpha$ spectra scanning the filament at 08:36 UT is impressive (see the linked animation). In Figure~\ref{rain}, panels a, b, c, and d show snapshots of the movie; the spectra 73, 75, 78, and 80 cross the north end of the filament between $y = 900 ~\mathrm{and}~ 1500$ pixels. At these locations, the filament 
shows a finger pattern (Figure~\ref{rain} panel f). The fingers are very faint, corresponding to a less dense plasma than in the filament.

The H$\alpha$ line exhibits an additional extension, resembling an elongated line of 0.5 \AA\ around 6565 \AA. This extension is seen as a fine structure in spectrum 73 at $y = 1655$ pixels (Figure~\ref{rain} panel a), slightly larger (covering a few more $y$ pixels) in the adjacent spectra 75 to 80 (Figure~\ref{rain} panels b, c, d).
Besides, we note that such high flow in the red wing of H$\alpha$  is observed in several consecutive spectra covering $x = 73-80$  $\approx$ 7$''$ during the raster (see panels a-d). It indicates that an extended plasma structure (around 7$''$ long) is falling. 
Moreover, the wavelength extension is connected to the H$\alpha$ line centre by an inclined line segment  (white arrow in panel d). The inclined structure suggests a progressive increase in the LOS velocity from $y \approx 1500$ pixels to $y \approx 1655$ pixels.

The profiles at the maximum extension location ($y ~\mathrm{pixel} = 1655$) in the consecutive spectra (73-80) show two dips: one at the centre wavelength of H$\alpha$ at 6562.8 \AA\ and a second around 6565 \AA\ (panel g). We quantify the measured Doppler shifts as the line displacements. The first dip corresponds to the chromosphere nearly at rest; the second dip corresponds to high downflows. The estimated maximum downflow velocity is around 100 \kms. In Section~\ref{CM}, we compute flow speeds using a cloud-model method to obtain more precise values of the supersonic speed, and in Section \ref{plasmoid}, we measure the spatial dimensions of the falling structure using images.

\subsection{Doppler shifts in the filament threads}
We used the second THEMIS observation at 08:37 UT, with its FOV centred on the northern part of the filament, which has higher spatial resolution. This observation is taken with a finer step size of 0.5$''$, not 1$''$; therefore, we have 161 spectra (slit positions) covering a FOV of 80$''$ in $x$. We chose this time because the counter-streaming flows in the main filament body and fingers in the northern part are clearly visible, unlike at other times.

In Figure~\ref{Doppler28}, spectra that cross the filament at two different locations are shown. In the top panels (panels a, b and c), the spectra cross the north end of the filament, which consists of many fine, finger-like strands.
The finger patterns are similar to the finger structures observed one minute earlier at 08:36 UT in Figure \ref{rain}, panel f. In Figure~\ref{Doppler28} panel b, the location of each finger is indicated by the cyan arrows. The H$\alpha$ line is wide at these positions. 


We used the technique described in \citet{karki2025} to obtain precise estimates of the filament strand size.
Cuts of the spectra are performed at three different wavelengths: H$\alpha$ line centre and H$\alpha$ $\pm$ 0.44 \AA, corresponding to the red and blue wings (yellow, red, and blue curves in Figure~\ref{Doppler28}c). The yellow curve obtained at the H$\alpha$ line centre shows many dips between $y=1000$ and $y=1600$ pixels; these dips correspond to the filament fingers. The horizontal cyan arrows from panel a to panel b indicate the position of each finger and correspond to the vertical arrows in panel (c). At the positions of the cyan arrows in panel c, in the yellow dips (the centres of the fingers), the red and blue curves also show a dip at the same location, indicating that the profile is broad. 
This can be due to thermal turbulent mechanisms or the integration of many structures along the LOS. These structures belong to the filament, with some possible contribution from the underlying chromosphere. 

In summary, the fingers observed in the line centre are also observed as thread-like extensions in the blue and red wings. Finally, in these fingers, at the end of the filament, no obvious counter-streaming flows are observed. At the location of the fingers the red wing is more extended than the blue wing (Figure~\ref{Doppler28}b), the absorption in the red wing is deeper than in the blue wing (Figure~\ref{Doppler28}c). This indicates that the fingers are red-shifted.

\subsection{Counter-streaming}
We compare the spectra of the fingers with those of the filament's main body in Figure~\ref{Doppler28}. The bottom panels of Figure~\ref{Doppler28} (panels d, e, f) are identical to the top panels but refer to the main body of the filament. The bottom panel d shows the intensity map of the filament at 08:37 UT, and the vertical black dashed line indicates the position of the spectra shown in panel e. The double black arrow indicates the filament. In this part of the filament, we observe counter-streaming flows, as follows. In the cut diagram (panel f), the blue and the red dips in the cuts obtained at $\pm$ 0.44 \AA\ are alternately placed, indicating a counter-streaming flow with a velocity greater than 20 \kms. The filament is dense enough to completely overlay the chromosphere. The high spatial resolution of the THEMIS spectra revealed counter-streaming within the filament's main core. The velocities of these flows are supersonic and nearly one order of magnitude greater than those of the counter-streaming flows discovered by \citet{Zirker1998}. These LOS counter-streaming flows had already been detected in filament spectra with such large velocity amplitudes \citep{karki2025}. The authors discussed various mechanisms that could produce such large alternating Doppler shifts between adjacent threads and concluded that they could be due to kink transverse waves. \cite{Goossens2014} showed that kink waves in spicules and mottles involve not only transverse motions but also rotational motions of solar magnetic flux tubes; the velocity field could be a spatially and temporally varying sum of both transverse and rotational motions. In \citet{karki2025}, no solution is convincing. The FWHMs of the  H$\alpha$ profiles of the filament are relatively large, about 1 \AA\ (Figure \ref{rain}g). This indicates either high micro-turbulence or a combination of many profiles corresponding to multiple structures along the LOS, each with different velocities. Therefore, multiple-thread radiative transfer models could better estimate the magnitude of these supersonic flows \citep{Gunar2008}.

\begin{figure}    
\centerline{\includegraphics[width=1.\textwidth,clip=]{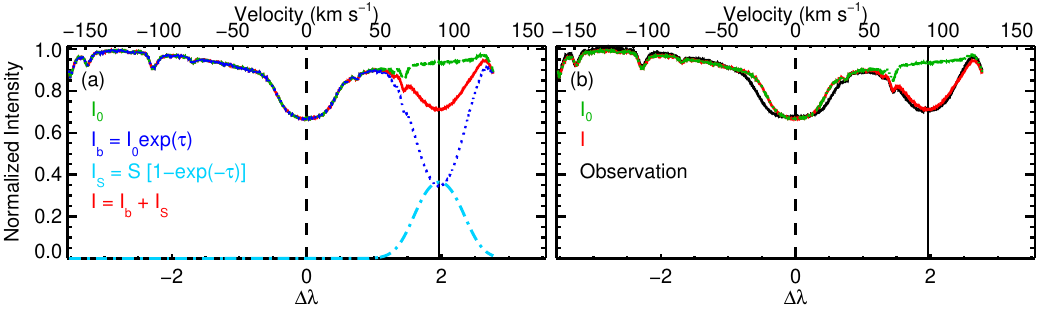}}
    \caption{The cloud model calculation of the H$\alpha$ profile observed by THEMIS at 08:36 UT for y = 1644 pixels and for slit position 78. The observed profile is covered by 2048 pixels in wavelength.
     The set of parameters corresponds to case 4 in Table \ref{tab:cloud_model}. The slit position 78 is shown as a green dashed line in panel f, and the corresponding spectra are shown in panel c of Figure~\ref{rain}. Panel a shows the cloud-model components, and panel b compares the observed profile (black line) with the cloud-model profile (red line). In panels a and b, the green line represents I$_0$, the background intensity profile  (reference profile) obtained in the chromosphere near the filament. The position of the reference profile along the slit is shown with a dashed black horizontal line in Figure~\ref{rain}c. The dotted blue line depicts $I_b = I_0\, \mathrm{{\bf e}}^{-\tau}$, the background intensity attenuated by the cloud. The dot-dashed cyan line shows $I_S = S\, [1 - \mathrm{{\bf e}}^{-\tau}]$, the contribution from the cloud. The solid red line shows $I= I_b + I_S$, the modelled intensity profile.}
\label{cloudmodel} 
\end{figure}

\subsection{Cloud model method}
\label{CM}
We model the observed H$\alpha$ profile of the downflow using the cloud-model technique. This method was first introduced by \cite{Beckers1964} and has since been widely applied to investigate asymmetric chromospheric line profiles using the H$\alpha$ line \citep{Mein1985, Mein1988, Schmieder2014, Verma2020} and the Mg~II lines \citep{Tei2018, Tei2020, Joshi2021}. This technique suits cold structures with high radial velocities above the chromosphere along the LOS and involves determining the background spectral profile. In this method, asymmetric line profiles are modelled using a plasma cloud. This provides evidence of plasma moving above the chromosphere.  
This method models the observed intensity $I(\Delta \lambda)$, where $\Delta \lambda = \lambda - \lambda_0$ is the wavelength difference between $\lambda$ and the central rest wavelength $\lambda_0$ of the H$\alpha$ line. 
The total intensity $I(\Delta \lambda)$ is given by the following relation when a single cloud is present in front of a background atmosphere with intensity $I_0$($\Delta \lambda$): 
\begin{equation}
    I(\Delta \lambda) = I_0(\Delta \lambda)\, \mathrm{e}^{-\tau(\Delta \lambda)} + S\,[1-\mathrm{e}^{-\tau(\Delta \lambda)}]
\end{equation}
   where $S$ is the source function, which is supposed to be uniform in the moving structure, and 
\begin{equation}
    \tau(\Delta \lambda) \equiv \tau_0\, \mathrm{exp} \left [-  \left (\frac{\Delta \lambda - \Delta \lambda_{LOS}}{\Delta \lambda_D} \right )^2 \right]
\end{equation}
    is the cloud optical thickness.
$\Delta \lambda_{LOS} = \lambda_0 V_{LOS}/c$ is the wavelength shift of a cloud with LOS velocity $V_{LOS}$.  
The Doppler width is
\begin{equation}
    \Delta \lambda_{D} \equiv \frac{\lambda_0}{c} \sqrt{\frac{2k_B T}{m_{H}} + V_{turb}^2}
    \end{equation}
where $T$ and $V_{turb}$ are the temperature and turbulent velocity of the cloud, respectively; $k_B$ is Boltzmann's constant; $m_H$ is the atomic mass of hydrogen; and $c$ is the speed of light. 

This method is robust for computing clouds with large Doppler shifts \citep{Mein1988}. To compare the observed data with the modelled data, we have  manually selected the parameters for the above equations in a classical  parameter space  \citep{Tei2018} with the following ranges:
\begin{itemize} 
\item the ratio of $S/I_0$ is between 0.5 and 0.9,  
\item the temperature between 5000 K and 10000 K,
\item the $\tau_0$ value between 0.5 and 1.23.
\item  the micro turbulence between 13.5 \kms\ and 18 \kms.
\end{itemize}

In our case, the H$\alpha$ profile we considered has two dips (e.g., Figure~\ref{rain}, panel g): one corresponding to the chromospheric rest profile and the other to the cloud. Since the dips in wavelength are well separated, the region of the second dip, $I(\Delta \lambda)$, is not affected by the chromospheric H$\alpha$ profile, so it is quite independent of the precise location used to define $I_0(\Delta \lambda)$. We used the reference profile defined near the filament to fit the first dip of the chromosphere and a cloud, defined by its source function S and optical thickness $\tau$, to fit the second dip. The fitting was quite good across all these solution sets (see Table~\ref{tab:cloud_model} in the Appendix).

Figure~\ref{cloudmodel} represents an  example of good fitting of the observed intensity $I(\Delta \lambda)$ and its modelled components using the following parameters:
\begin{enumerate}
\item Turbulent velocity $V_{turb}$: 17.5 \kms,
\item The optical thickness $\tau_0$: 0.99,
\item The ratio of the source function S of the cloud to the background intensity $I_0$ at the line centre of H$\alpha$ $(S/I_0)$: 0.84
\end{enumerate}
The ratio $(S/I_0)$ varies across parameter sets but remains close to 1. 
Using the parameter sets in the range mentioned above, we found a mean Doppler shift of 89.8 \kms\ with a standard deviation of 0.2 \kms\ (calculated from values obtained with different parameter sets). This measured velocity represents the statistical dispersion of the velocities obtained from different model fits. 

The vertical velocity component can exceed the Doppler shift due to the filament's location and perspective effects. However, the filament is near the disk centre, so this effect is small.  This Doppler velocity is comparable to the free-fall velocity of about 100 \kms\ if we assume the downflowing plasma is initially at rest at a height of 20 Mm above the chromosphere, a value derived from a 3D MHD reconstruction of the filament height in a previous paper \citep{Karki2026}. 

With the cloud model method, Doppler velocities have been well estimated since \cite{Mein1988}. The other parameters are strongly coupled, making it difficult to derive precise values for the source function (a), optical thickness $\tau$, and, consequently, density and mass. Due to the coupling of all these parameters, there are no linear evolution trends between a, $\tau$, and the results, as we see in Table~2. Calibrating the H$\alpha$ profiles and using a radiative transfer model to derive these quantities would be useful; however, this is outside the scope of this paper.

Next, we estimate the different uncertainties existing in the observation profiles and see how they propagate in the results.  Spectrograph wavelength calibration introduces uncertainty. The diffraction grating provides a dispersion per meter (or pixel) on the CCD camera. This must be converted to wavelength, introducing uncertainty. \cite{Peat2026} studied the calibration of the THEMIS data for the same campaign of observations in September 2023. They fitted the observed mean  H$\alpha$ profile  with the reference profile of  the solar atlas  of \citet{Delbouille1973}. A conservative uncertainty in the
wavelength calibration of approximately 40 m\AA\ was found: + 40 m\AA\ in the  extended blue wing and  - 40 m\AA\ in the extended red wing. This implies an uncertainty of $\pm$ 1.8 \kms\ on the LOS velocity.

In this study we employed a different method to calibrate the wavelength, we used the telluric lines, which have no Dopplershifts. These  telluric lines lead to estimated velocities in the far wings of H$\alpha$ with an error of one pixel, and a slight blue or red shift in the line center. 
Moreover, the wavelength uncertainty in the observed profiles (intensity versus $\lambda$) is weak because of THEMIS's high spectral dispersion (3.067 m\AA\ per pixel).  
Therefore, the wavelength shift of the cloud with the LOS velocity ($V_{LOS}$), $\Delta \lambda_{LOS} = \lambda_0 V_{LOS}/c$,  
is also determined with a small uncertainty (pixel size = $\sim$ 0.15 \kms).

A next systematic uncertainty is in the computation of $\Delta \lambda = \lambda - \lambda_0$ since the central rest wavelength $\lambda_0$ of the H$\alpha$ line cannot be provided by atomic data. The CCD measurement is in pixels (spatial length), so it is not related to atomic data. The typical way of previous studies to estimate $\lambda_0$ is with the mean profile taken in a large quiet region, so without extra velocities. This could introduce bias.  With the present data, we can use a more direct method than typically used previously because of the large plasma velocity present, so that the corresponding spectral profile is shifted away from the rest profile (see an example in Figure \ref{rain} panels c, g). 
We select in our spectra narrow profiles of the underlying chromosphere, which is nearly at rest, so with no positive or negative extensions of the spectral profile. 
The observed profiles show a large separation with two different dips; one corresponds to the blob, and the other to the underlying chromosphere (Figure~\ref{cloudmodel} black curve in panel b).  First, the effect of the chromosphere is not entangled in the observed spectra as in the case of observing a filament.  In our case, radiative transfer is simple because we obtain the spectra of the chromosphere and of the moving plasma separately.  So the Doppler shifts of the blobs are independent of the chromospheric profile. 

As a side effect, this method also corrects for the solar rotation.
The LOS velocity of the solar rotation is slightly changing in the whole observed FOV, especially when going closer to the limb. In our case, the observations are nearly at the centre of the disk, and the reference profile is at a small distance from the concerned area.

We finally cumulate the above uncertainties supposing statistical independence (so summing up the variances).  The total uncertainty on the estimated velocity is $\sim$ 0.25 \kms\ so well lower than the deduced plasma velocity ($\sim$ 90 \kms). This uncertainty is also much smaller than the added micro-turbulence included in the model to fit the spectral line width ($\sim$ 18 \kms). This represents differential velocity within the down flowing plasma, and this physical dispersion of the velocity largely overpass the measurement uncertainties.  
\begin{figure}    
\centerline{\includegraphics[width=1.\textwidth,clip=]{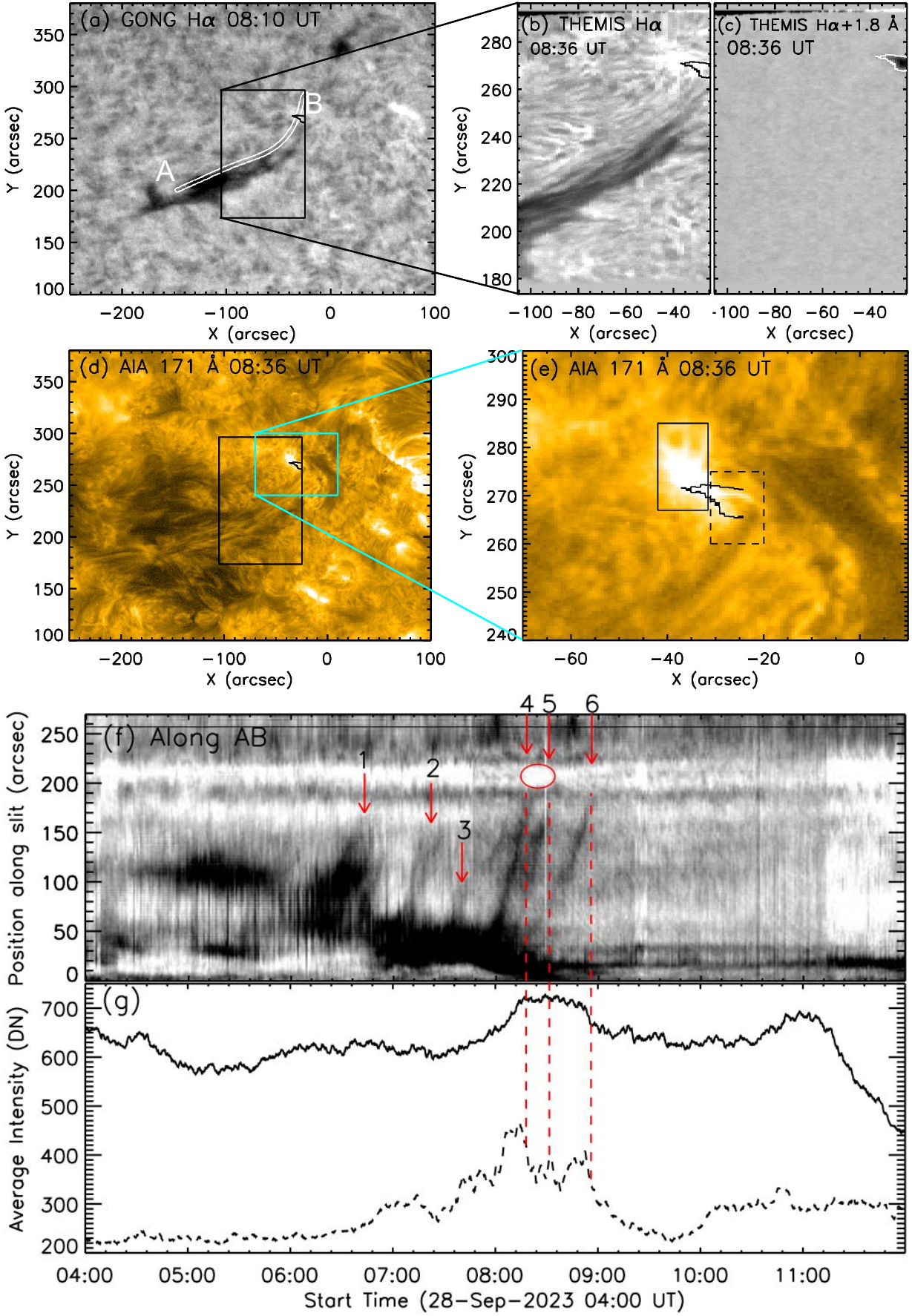}}
       \caption{Filament and BP region observed with GONG (panel a), THEMIS  at the H$\alpha$ centre (panel b), at H$\alpha$ centre + 1.8 \AA\ (panel c), and AIA filter 171 \AA\ (panels d, e). In panel (a), the white curve AB is the artificial slit used to detect plasma motions along the filament shown in the time-distance diagram (panel f). The black box in panels a and d indicates the THEMIS FOV, depicted in panels b and c. The contour of the dark region of panel c is overlaid in panels a, b, d, and e. Panel (e) shows a zoomed view of panel (d) centred on the BP. 
       In panel f, the red arrows 1-6 indicate repetitive plasma motions along the filament. The enhanced H$\alpha$ brightness is denoted by a red ellipse. The red arrows indicate the plasma arrival times, marked by the red dashed straight lines labelled 4, 5, and 6, corresponding to 08:18:21 UT, 08:31:57 UT, and 08:56:21 UT, respectively. Panel g represents the AIA 171 \AA\ intensity variations in the two boxes drawn in panel e. The light curve peaks at 08:13:57 UT, 08:32:09 UT, and 08:52:45 UT. An animation of panels a and e is attached to this figure.}
\label{gong_aia}
\end{figure}

\begin{figure}    
\centerline{\includegraphics[width=1.\textwidth,clip=]{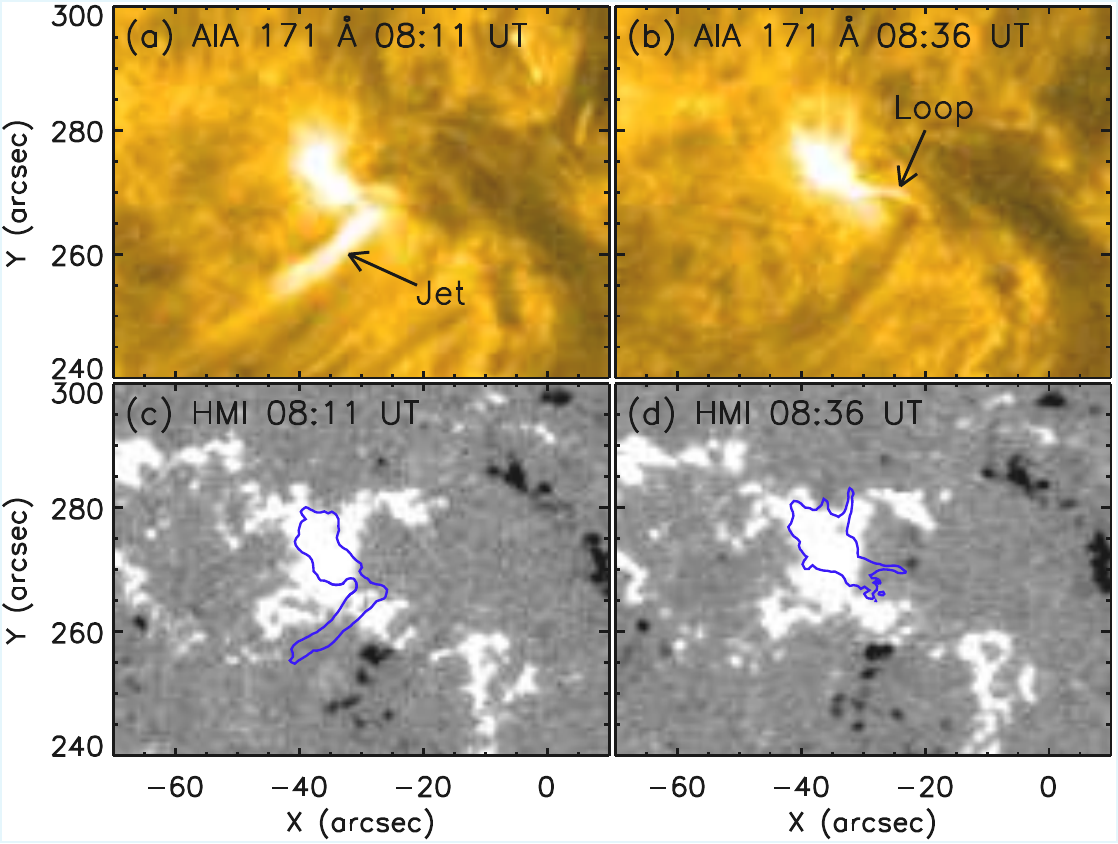}}
\caption{Zoomed view of the BP and jet in AIA 171 \AA\ channel (panels a and b), the corresponding HMI magnetograms saturated at levels $\pm 80$ G (panels c and d). The blue contours in panels c and d are the BP and jet contours observed in the AIA 171 \AA\ channel. } 
\label{SDO}
\end{figure}

\section{Flows along the filament from GONG observations}
\label{plasmoid}
THEMIS observation mode does not allow a series of observations to study plasma motion along the filament (Table~\ref{tab:THEMIS}). Therefore, we used GONG observations to track the filament's evolution. In the GONG H$\alpha$ movie attached to Figure~\ref{gong_aia}, the observed filament undergoes significant changes throughout the day. To trace plasma motion along the filament spine, we commonly use time-distance analysis with a curved slit. We did this in a previous paper for September 28 and 29 \citep{Karki2026}. Large-amplitude oscillations have been detected in the filament's main body.

However, the filament's northern end exhibits different behaviour from the main body, as the dark moving plasma moves repetitively without returning (see the GONG movie attached to Figure~\ref{gong_aia}). 
The north end of the filament corresponds to the THEMIS observations with supersonic high downflow (Figure~\ref{gong_aia} panels b, c). 
In panel c and in the movie of the THEMIS images, the structure observed in the far H$\alpha$ red wing consists of an elongated thread and a blob moving from south to north towards the bright point, with increasing speed. The elongated thread is on the order of 10$''$ in length and a few pixels in width (maximum 1$''$), while the trailing part of the observed blob is more rounded, with a size on the order of 4.3$''$. These dimensions are consistent with the histogram of coronal rain blobs \citep{Antolin2012}.

We examined the plasma flows along these filament structures from 04:00 UT to 12:00 UT on September 28, 2023, and, for this purpose, we aligned the GONG H$\alpha$ data to a common reference time (09:00 UT on September 28, 2023). We selected a slit AB parallel to the filament but slightly north (Figure~\ref{gong_aia} panel a). The slit spans from the southeast (point A) to the northwest (point B) of the filament. Figure~\ref{gong_aia} (panel f) shows the time-distance plot along the slit AB. 
The time-distance plot reveals a repetitive pattern of straight, parallel, darker bands (labelled from 1 to 6). It is not an oscillation pattern since the plasma is not returning. This means the plasma moves repetitively at 60--80 \kms. The moving plasma blob also elongates over time (see GONG movie).

According to the 3D reconstruction of the filament \citep{Karki2026}, the plasma could be located in multiple dips in horizontal magnetic field lines anchored in the photosphere. Any photospheric motion would induce plasma motion in the dips, potentially leading to escape along bent magnetic field lines.

We interpret this evolution as plasma motions along the filament's magnetic field lines bending down towards the chromosphere. Additionally, at the slit location $\approx 210''$, the background is slightly brighter (Figure~\ref{gong_aia} panel f). Furthermore, at the arrival times of the dark bands around the slit location ($\approx 190''$), the brightening region is further enhanced. The brightening corresponding to bands 4-5 is indicated by a red circle in panel f of Figure~\ref{gong_aia}. The next section analyses the relationship between plasma motion and enhanced brightness.

\section{EUV observations}
\label{EUV}
\subsection{Bright area and filament channel}
The darker part of the filament observed in AIA 193, 171, 211, and 304 \AA\ (Figure~\ref{AIA_BP}) corresponds to the filament observed in H$\alpha$. We note the presence of a BP at the northeast end of the filament across all AIA hot EUV channels, indicated by white arrows in Figure~\ref{AIA_BP}{b--e}. The BP is visible only in the hot EUV channels, not in the 304 \AA\ channel. This implies no chromospheric heating; the plasma is heated to coronal temperatures. Moreover, the plasma blobs moving toward the northern end of the filament do not emit in EUV. We focus on the evolution of this BP (see the movie linked to Figure~\ref{AIA_BP} and \ref{gong_aia}). Moreover, the brightness of this small area mainly coincides with the positive magnetic polarity of the network at the intersection of supergranules in the HMI maps (Figure~\ref{SDO}). The positive magnetic flux measurements (2.7 $\times$ $10^{20}$ Mx) at the BP location in HMI maps are two orders of magnitude larger than the negative flux associated with the nearby negative patches. The evolution of the two fluxes does not appear related. Finally, on September 29, the BP no longer exists and is replaced by a curled filament \citep{Karki2026}.

\subsection{Light curves of the bright point}
\label{lc}
On September 28, we divided the BP brightness region into two smaller regions (boxes in Figure~\ref{gong_aia}, panel e) because they evolved differently. The solid black box is centred on the network brightness; the dashed black box corresponds to the region of the supersonic flows outlined by the black contour (see panel c for its definition). The uncertainty in the alignment between THEMIS and GONG is estimated at 1$''$. A similar uncertainty also applies to AIA 171 \AA. This alignment uncertainty is well lower than the blob's length and width (10$''$ $\times$ 4$''$). 

An example of the light curves at 171 \AA\ 
is shown in Figure~\ref{gong_aia} (panel g), using the same drawing convention as the boxes shown in panel (e). Both light curves peak between 07:00 and 09:00 UT. The temporal distribution of the peak maxima nearly corresponds to the arrival times of the plasma near the BP (darker bands in Figure~\ref{gong_aia} panel f). To be more precise, we note that the brightness peaks between 08:00 UT and 09:00 UT occur 1 to 5 minutes before the lines marked with dashed red vertical lines (lines 4, 5, 6 in Figure~\ref{gong_aia} panel g). This may indicate that the blobs were slightly heated beforehand.
Next, the timescale between EUV peaks is not truly periodic. Moreover, in the GONG movie, we do not see the plasma return after the moving plasma, so the plasma motion does not exhibit oscillatory characteristics.

Snapshots from the BP zoom movie are shown in Figure~\ref{SDO}, corresponding to the two peaks in brightness (Figure~\ref{gong_aia}, panel g). The first peak (line 4) coincides with a bright jet in AIA 171. This hot jet is ejected from the BP in the filament channel (Figure~\ref{SDO}a). The second peak (line 5) corresponds to a bright loop. Finally, the third peak, near line 6, has just a brighter area (not shown). 

The zoom of the BP in the dashed box at 08:36 UT shows a mini loop (Figure~\ref{SDO}b). This loop-like structure first appears at 08:35 UT in AIA 171 \AA, then brightens, reaches maximum intensity at 08:37 UT, and disappears at 08:39 UT. The plasma flowing from the filament could hit the loop at 08:36 UT (see the attached movie related to Figure~\ref{gong_aia}). This relationship between the high downflows and the BP brightness maxima, in both time and location, suggests an interaction between the filament and the coronal magnetic field lines leading to strong localised heating. The high temperature detected in the BP visible in AIA 94 \AA\ (6 MK) suggests that magnetic reconnection could be located in the corona, since there is no enhancement in AIA 304 and it is very weak in H$\alpha$.

\subsection{Coronal rain scenario}

In this part, we investigate whether the coronal rain mechanism could be the origin 
of the supersonic downflows, as follows. In the review by \citet{Keppens2025}, many theoretical possibilities for obtaining cool plasma from the hot corona are discussed, applicable to prominence formation, coronal rain, or post-flare loops. Thermal instability can occur in hot plasma (coronal), producing cold plasma (chromospheric-like) that flows downward in the quiet Sun and during flares. Coronal rain is also present during reconnection-driven flares, and condensed plasma flows downward at high speeds along post-flare loops \citep{Malherbe1997}.
In our case, the cold downflowing plasma would originate from an earlier thermal instability (creating cold plasma), which is not captured in dips, so it does not build up the filament but instead flows along the field lines \citep{Keppens2025}. So, downflowing plasma would be linked to filament formation by thermal instability of the coronal plasma. 

In quiet regions, coronal rain is very rare in disk observations, so this would be the first time we could measure Doppler shifts of condensed plasma, a signature of coronal rain \citep{Antolin2012, Froment2020}. In \cite{Froment2020}, they estimated the velocity and density of the rain blobs using a Gaussian method and the differential emission measure (DEM) computed from the AIA filters, respectively. They also detected periodic pulsations of the thermal non-equilibrium instability with a 6-hour period. Their velocities are close to our estimates.
We also detect long-term oscillations of 4-5 hours in the BP in AIA 171. However, we are not in a quiet region because the BP is close to a filament channel in a network region (with a magnetic field strength of about 160--210 G). Finally, coronal rain would be easier to detect in a region without a filament channel.

\section{Discussion and Conclusion}
\label{Conclusion}
Coordinated observations have been obtained during a THEMIS-IRIS campaign on September 28, 2023. THEMIS detected multiple strands in the filament spine and at one of its ends. The high spatial resolution of the instrument, with adaptive optics, allows us to confirm the existence of LOS counter-streaming in the filament main body, on the order of 20 \kms\ between adjacent threads, each with an arcsecond diameter, as found in \citet{karki2025}. In the previous paper, no real explanation was found. Counter-streaming flows may be relevant to kink or torsional waves, as seen in mottles \citep{Goossens2014}, but this has not yet been validated for filaments. We should explore radiative-transfer filament models with multiple threads in the future \citep{Gunar2008}.

THEMIS mode MTR provides high-spatial and -spectral resolution spectra across a wide wavelength domain in H$\alpha$ ($\pm$ 3 \AA\ or $\pm$ 150 \kms) with an uncertainty of 0.15 \kms\ (spectral resolution of 3 m\AA) as it scans the filament environment. This capability allows us to detect supersonic downflows in H$\alpha$, using the cloud model method,  with Doppler shifts equal to  89.8 \kms\ and a standard dispersion of $\pm$ 0.2 \kms, of the same order as the uncertainty of the measurement of the velocities. Next, we measure plasma flows using time-distance analysis of GONG H$\alpha$ intensity data. The plasma moves along the filament toward its northern end, where H$\alpha$ downflows are observed. The speed orthogonal to the LOS is on the order of 80 \kms. Finally, the plasma is not oscillating. 

From THEMIS, GONG, and AIA observations, we suggest the following scenario: plasma moving along the filament strands, away from the main dips, could fall toward the chromosphere. A ballistic approach yields consistent results with this scenario, as follows. Using the filament height (20 Mm), derived from a 3D MHD reconstruction \citep{Karki2026} we compute the free-fall speed of blobs reaching the chromosphere to be around 100 \kms, and the time to do so around 6 min, which is comparable to observations. The main observational constraints on the physical mechanism are that no extra heating is detected in AIA 304 \AA\ and that the H$\alpha$ intensity enhancement is weak.

This cool, downflowing plasma occurs near a BP visible in AIA EUV.  It also coincides with brightening across all channels from 94 \AA\ to 171 \AA, except at 304 \AA.
Therefore, a second possible mechanism is a thermal instability that occurs after a magnetic reconnection episode and leads to coronal rain. Still, the timing does not inform us about the causality between the observed downflows and the enhanced EUV emission, and we do not have long enough THEMIS observation sequences to conclude.

For coronal rain due to magnetic reconnection, we need EUV spectropolarimetry to diagnose magnetism from the solar photosphere to the transition region and to explore how the magnetic field evolves from the dynamically driven photosphere to the magnetically dominated corona, where twisted non-potential flux ropes form. It was the concept of a future NASA instrument \citep[CMEx,][]{Gilbert2013}.


\appendix   
Table \ref{tab:THEMIS} summarises the observations obtained by THEMIS/MTR on September 28, 2023, including the $x$-step size and the corresponding FOV covered. Table \ref{tab:cloud_model} presents the results of the cloud model fitting for different sets of parameters. We fit the model profiles to the observed profile by varying the parameters within a standard parameter space. We determine the best-fit model by minimising the root-mean-square (RMS) difference between the observed and modelled profiles (see Table~\ref{tab:cloud_model}). RMS values are computed using $RMS = \sqrt{mean~(observation - model)^2}$, where the observation and model represent the observed and modelled profiles, respectively, both normalised by the maximum value of $I_0$. For a different set of parameters, the RMS value is almost similar around 0.015, as mentioned in Table \ref{tab:cloud_model}.

\begin{table*}[!ht]
    \begin{tabular}{ccccc}
    \hline
    S. No.& File & Time (UT) & x step (arcsec) & FOV  \\
        \hline
       1. & t01 & 08:23 & 1 & Center (F$_2$) \\
       2. & t03 & 08:32 & 1 & Center (F$_2$)\\
       3. & t04 & 08:36 & 1 &  West\\
       4. & t05 & 08:37 & 0.5 & West \\
       5. & t06 & 08:39 & 0.5 & West \\
       6. & t08 & 08:45 & 0.5 & West\\
       7. & t09 & 08:48 & 0.5 & West\\
       8. & t10 & 08:54 & 0.5 & Center (F$_2$)\\
       9. & t13 & 09:50 & 0.5 & Center (F$_2$)\\
      10. & t14 & 10:10 & 0.5 & East (F$_1$) \\
      11. & t16 & 10:38 & 0.5 & Center (F$_2$) \\
      12. & t19 & 11:10 & 0.5 & Center (F$_2$) \\
      13. & t20 & 11:19 & 0.5 & Center (F$_2$) \\
      \hline
    \end{tabular}
    \caption{List of observations taken by THEMIS/MTR on September 28  2023, along with the x-step size and the FOV covered.}
    \label{tab:THEMIS}
\end{table*}

\begin{table*}[!h]
    \begin{tabular}{ccccccc}
    \hline
    S. No.  & Temperature (K) & S/I$_0$ & V$_{LOS}$ (\kms)  & V$_{turb}$ (\kms)  & $\tau_0$  & RMS \\
    \hline
      1.  & 5000 & 0.50 & 90.0  & 18.0 & 0.51 & 0.0156\\
      2.  & 5000 & 0.80 & 89.5  & 17.0 & 0.90 & 0.0153\\
      3.  & 5000 & 0.90 & 89.8  & 16.0 & 1.23 & 0.0154\\
      4.  & 5000 & 0.84 & 89.5  & 17.5 & 0.99 & 0.0151\\
      5.  & 7000 & 0.50 & 90.0  & 17.0 & 0.51 & 0.0156\\
      6.  & 7000 & 0.80 & 90.0  & 16.0 & 0.89 & 0.0156\\
      7.  & 7000 & 0.90 & 89.5  & 15.5 & 1.22 & 0.0151\\
      8.  & 10000 & 0.50 & 90.0  & 16.0 & 0.51 & 0.0155\\
      9.  & 10000 & 0.80 & 89.8  & 14.5 & 0.89 & 0.0155\\
      10. & 10000 & 0.90 & 89.8  & 13.5 & 1.20 & 0.0155\\
      \hline
    \end{tabular}
    \caption{Results of the cloud model for a different set of parameters.}
    \label{tab:cloud_model}
\end{table*}

\begin{acks}
We would like to acknowledge the reviewer for the valuable comments and suggestions, which have improved the manuscript. This work is based on ground-based observations obtained by the THEMIS telescope in Tenerife in the Canary Islands, operated by Bernard Gelly, Richard Douet, and Didier Laforgue during a multi-wavelength campaign with IRIS (IHOP444 - PIs Nicolas Labrosse - Brigitte Schmieder). We sincerely thank the entire THEMIS team for these observations. AIA data are courtesy of NASA/SDO and the AIA, EVE, and HMI science teams. IRIS is a NASA small explorer mission developed and operated by LMSAL, with mission operations executed at NASA Ames Research Centre and major contributions to downlink communications funded by ESA and the Norwegian Space Centre. B.S.\ thanks Rony Keppens and Tom Van Doorsselaere for fruitful discussions on waves and filament fine structures. S.P.\ is funded by the European Union (ERC-AdG agreement No 101141362, Open SESAME). Views and opinions expressed are, however, those of the author(s) only and do not necessarily reflect those of the European Union or the European Research Council. Neither the European Union nor the granting authority can be held responsible. S.P.\ is also funded via the projects C16/24/010 (C1 project Internal Funds KU Leuven), G0B5823N and G002523N (WEAVE) (FWO-Vlaanderen), and 4000145223 (SIDC Data Exploitation (SIDEX2), ESA Prodex). LUNEX EuroMoonMars company for supporting her mission to Trivandrum (INDIA) where this paper was discussed.
\end{acks}

\begin{authorcontribution}
GK performed the formal data analysis and validation. GK and P.\ Devi edited the manuscript. BS and RC supervised the work and contributed to the manuscript's writing. SP and P.\ D\'emoulin revised the manuscript. 
\end{authorcontribution}

\begin{fundinginformation}
G.K.\ acknowledges the Department of Science and Technology, New Delhi, India, for the INSPIRE fellowship. R. C. acknowledges the support from ANRF project No. EEQ/2023/000214. P. Devi\ acknowledges the support from the Centres of Excellence scheme of the Research Council of Norway, project number 262622. S.P.\ is funded by the European Union (ERCEA, ERC-AdG agreement No 101141362, Open SESAME). Views and opinions expressed are, however, those of the author(s) only and do not necessarily reflect those of the European Union or the European Research Council. Neither the European Union nor the granting authority can be held responsible. These results were also obtained in the framework of the projects C16/24/010 (C1 project Internal Funds KU Leuven), G0B5823N and G002523N (WEAVE) (FWO-Vlaanderen), and 4000145223 (SIDC Data Exploitation (SIDEX2), ESA Prodex). BS thanks  LUNEX EuroMoonMars for supporting her mission to Trivandrum (INDIA) where this paper was discussed.
\end{fundinginformation}

\begin{dataavailability}
The datasets used in the present study are available at \url{http://jsoc.stanford.edu/}, \url{https://iris.lmsal.com/search/}, \url{https://gong.nso.edu/}. 
\end{dataavailability}



\begin{ethics}
\begin{conflict}
The authors declare no conflicts of interest.
\end{conflict}
\end{ethics}


\bibliographystyle{spr-mp-sola}
\bibliography{references}  


\end{document}